\documentclass[12pt]{article}
\usepackage{amssymb,latexsym}
\usepackage{epic,eepic}

\newcommand{\2}{\vspace{0.5 cm}}

\newcommand{\qed}{\hfill$\Box$}
\newcommand{\pf}{{\bf Proof: }}
\newtheorem{theorem}{Theorem}[section]

\newtheorem{proposition}[theorem]{Proposition}
\newtheorem{lemma}[theorem]{Lemma}

\newtheorem{corollary}[theorem]{Corollary}
\newtheorem{conjecture}[theorem]{Conjecture}

\newcommand{\beq}{\begin{equation}}
\newcommand{\eeq}{\end{equation}}

\begin{document}

\title{Note on edge-colored graphs and digraphs without properly colored cycles}

\date{}

\author{Gregory Gutin\\ Department of Computer Science\\ Royal Holloway, University
of London\\ Egham, Surrey, TW20 0EX, UK\\ Gutin@cs.rhul.ac.uk}
\maketitle

\begin{abstract}
We study the following two functions: $d(n,c)$ and $\vec{d}(n,c)$;
$d(n,c)$ ($\vec{d}(n,c)$) is the minimum number $k$ such that every
$c$-edge-colored undirected (directed) graph of order $n$ and
minimum monochromatic degree (out-degree) at least $k$ has a
properly colored cycle. Abouelaoualim et al. (2007) stated a
conjecture which implies that $d(n,c)=1.$ Using a recursive
construction of $c$-edge-colored graphs with minimum monochromatic
degree $p$ and without properly colored cycles, we show that
$d(n,c)\ge {1 \over c}(\log_cn -\log_c\log_cn)$ and, thus,  the
conjecture does not hold. In particular, this inequality
significantly improves a lower bound on $\vec{d}(n,2)$ obtained by
Gutin, Sudakov and Yeo in 1998.

{\em Keywords: edge-colored graphs, properly colored cycles.}
\end{abstract}

\section{Introduction}

All directed and undirected graphs considered in this paper are
simple, i.e., have no loops or parallel edges. We consider only
directed cycles in digraphs; the term cycle (in a digraph) will
always mean a directed cycle.

Let $G=(V,E)$ be a directed or undirected graph, and let $\chi: E
\rightarrow \{1,2,\ldots,c\}$ be a fixed (not necessarily proper)
edge-coloring of $G$ with $c$ colors, $c \geq 2$. With given $\chi$,
$G$ is called a {\em $c$-edge-colored} (or, {\em edge-colored})
graph. A subgraph $H$ of $G$ is called {\em properly colored\/} if
$\chi$ defines a proper edge-coloring of $H$, i.e., no vertex of $H$
is incident to a pair of edges of the same color. For a vertex of a
$c$-edge-colored graph $G$,  $d_i(x)$ denotes the number of edges of
color $i$ incident with $x$. Let $\delta_{mon}(G)=\min\{d_i(x):\
x\in V(G),\ i\in \{1,2,\ldots ,c\}\}.$ If $G$ is directed,
$d^+_i(x)$ denotes the number of edges of color $i$ in which $x$ is
tail. Let $\delta^+_{mon}(G)=\min\{d^+_i(x):\ x\in V(G),\ i\in
\{1,2,\ldots ,c\}\}.$

The authors of \cite{abou2} stated the following:

\begin{conjecture}
Let $G$ be a $c$-edge-colored undirected graph of order $n$ with
$\delta_{mon}(G)=d\ge 1$. Then $G$ has a properly colored cycle of
length at least $\min\{n,cd\}$. Moreover, if $c>2$, then $G$ has a
properly colored cycle of length at least $\min\{n,cd+1\}$.
\end{conjecture}

In the next section, using a recursive construction of
$c$-edge-colored graphs with minimum monochromatic degree $d$ and
without properly colored cycles, we show that this conjecture does
not hold. Moreover, for every $d\ge 1$ there exists an edge-colored
graph $G$ with $\delta_{mon}(G)\ge d$ and with no properly colored
cycle.

We will study the following two functions: $d(n,c)$ and
$\vec{d}(n,c)$; $d(n,c)$ ($\vec{d}(n,c)$) is the minimum number $k$
such that every $c$-edge-colored graph (digraph) of order $n$ and
minimum monochromatic degree (out-degree) at least $k$ has a
properly colored cycle. Gutin, Sudakov and Yeo \cite{gutinDM191}
proved the following bounds for $\vec{d}(n,2)$

\begin{equation}\label{GSYbound}
{1 \over 4}\log_2 n + {1\over 8} \log_2\log_2 n + \Theta(1)\le
\vec{d}(n,2)\le \log_2 n -{1 \over 3} \log_2\log_2 n + \Theta(1)
\end{equation}

Using our construction, we prove that $\vec{d}(n,2)\ge {1 \over
2}(\log_2 n -\log_2\log_2n)$. This improves the lower bound in
(\ref{GSYbound}). (The lower bound in (\ref{GSYbound}) was obtained
using significantly more elaborate arguments.) This bound on
$\vec{d}(n,2)$ follows from lower and upper bounds on $d(n,c)$ and
$\vec{d}(n,c)$ obtained for each value of $c.$ The bounds imply that
$d(n,c)=\Theta(\log_2 n)$ and $\vec{d}(n,c)=\Theta(\log_2 n)$ for
each fixed $c\ge 2.$

Properly colored cycles have been studied in several papers, for a
survey, see Chapter 11 in \cite{bang2000}. Properly colored cycles
in 2-edge-colored undirected graphs generalize cycles in digraphs
and are of interest in genetics \cite{bang2000}. More recent papers
on proporly colored cycles include
\cite{abou1,abou2,fleischnerGC21}. Interestingly, the problem to
check whether an edge-colored undirected graph has a properly
colored cycle is polynomial time solvable (we can even find a
shortest properly colored cycle is polynomial time \cite{abou1}),
but the same problem for edge-colored digraphs is NP-complete
\cite{gutinDM191}.


\section{Results}

\begin{theorem} For each $d\ge 1$ there is an edge-colored graph $G$
with $\delta_{mon}(G)= d$ and with no properly colored cycle.
\end{theorem}
\pf Let $(p_1,p_2,\ldots ,p_c)$ be a vector with nonnegative
integral coordinates $p_i$.  For an arbitrary $(p_1,p_2,\ldots
,p_c)$, $G(p_1,p_2,\ldots ,p_c)$ is recursively defined as follows:
take a new vertex $x$ and graphs $H_1=G(p_1-1,p_2,p_3,\ldots
,p_{c-1},p_c)$ if $p_1>0$, $H_2=G(p_1,p_2-1,p_3,\ldots
,p_{c-1},p_c)$ if $p_2>0$, $\ldots$, $H_c=G(p_1,p_2,p_3,\ldots
,p_{c-1},p_c-1)$ if $p_c>0$ and add an edge of color $i$ between $x$
and and every vertex of $H_i$ for each $i$ for which $p_i>0.$ In
particular, $G(0,0,\ldots ,0)=K_1$.

It is easy to see, by induction on $p_1+p_2+\cdots +p_c,$ that
$G=G(p_1,p_2,\ldots ,p_c)$ has no properly colored cycle and
$\delta_{mon}(G)= \min\{p_i:\ i=1,2,\ldots ,c\}.$\qed

\2

In fact, for each $d\ge 1$ there are infinitely many edge-colored
graphs $G$ with $\delta_{mon}(G)= d$ and with no properly colored
cycle. Indeed, in the construction of $G(p_1,p_2,\ldots ,p_c)$ above
we may assume that $G(0,0,\ldots ,0)$ is an edgeless graph of
arbitrary order.

\begin{lemma}\label{Glemma}
Let $n(p_1,p_2,\ldots ,p_c)$ be the order of $G(p_1,p_2,\ldots
,p_c)$ and let $n_c(p)=n(p_1,\ldots ,p_c)$ for $p=p_1=\cdots =p_c.$
Then $n(p_1,\ldots ,p_c)\le s2^s$, where $s=p_1+p_2+\ldots + p_c$,
provided $s>0$ and $p\ge {1 \over c}(\log_cn_c(p)
-\log_c\log_cn_c(p)).$
\end{lemma}
\pf We first prove $n(p_1,\ldots ,p_c)\le s2^s$  by induction on
$s\ge 1.$ The inequality clearly holds for $s=1$. By induction
hypothesis, for $s\ge 2$, we have

\begin{eqnarray*}
n(p_1,\ldots ,p_c) & \le &  1+ \sum \{ n(p_1,\ldots
,p_{i-1},p_i-1,p_{i+1},\ldots ,p_c):\ p_i>0, i=1,2,\ldots ,c\}\\
& \le & 1+c(s-1)c^{s-1}  \le  sc^s
\end{eqnarray*}

Thus, $n_c(p)\le cp \cdot c^{cp}.$ Observe that $n_c(p)>ac^a$
provided $a=\log_cn_c(p) -\log_c\log_cn_c(p)$ and, thus, $cp\ge
\log_cn_c(p) -\log_c\log_cn_c(p).$\qed

\begin{corollary}
We have $\vec{d}(n,c)\ge d(n,c)\ge {1 \over c}(\log_cn
-\log_c\log_cn).$
\end{corollary}
\pf Let $H$ be a $c$-edge-colored undirected graph and $H^*$ be a
digraph obtained from $H$ by replacing every edge $e=xy$ with arcs
$xy$ and $yx$ both of color $\chi(e).$ Clearly, $H$ has a properly
colored cycle if and only if $H^*$ has a properly colored cycle.
Thus, $\vec{d}(n,c)\ge d(n,c)$. The inequality $d(n,c)\ge {1 \over
c}(\log_cn -\log_c\log_cn)$ follows from Lemma \ref{Glemma} and the
fact that graphs $G(p,p,\ldots ,p)$ have no properly colored
cycles.\qed

\2

We see that $\vec{d}(n,2)\ge {1 \over 2}(\log_2 n -\log_2\log_2n)$.
This is an improvement over the lower bound on $\vec{d}(n,2)$ in
(\ref{GSYbound}). Using the upper bound in (\ref{GSYbound}), we will
obtain an upper bound on $\vec{d}(n,c)$ and, thus, $d(n,c).$

\begin{proposition}
We have $\vec{d}(n,c)\le {1 \over {\lfloor c/2 \rfloor}} (\log_2 n
-{1 \over 3} \log_2\log_2 n + \Theta(1)).$
\end{proposition}
\pf Let $D$ be a $c$-edge-colored digraph of order $n$ with
$\delta_{mon}(D)\ge {1 \over {\lfloor c/2 \rfloor}} (\log_2 n -{1
\over 3} \log_2\log_2 n + \Theta(1)).$ Let $D'$ be the
2-edge-colored digraph obtained from $D$ by assigning color 1 to all
edges of $D$ of color $1,2,\ldots ,\lfloor c/2 \rfloor$ and color 2
to all edges of $D$ of color $\lfloor c/2 \rfloor + 1,\lfloor c/2
\rfloor + 2,\ldots, c.$ It remains to observe that
$\delta_{mon}(D')\ge \log_2 n -{1 \over 3} \log_2\log_2 n +
\Theta(1)$ and every property colored cycle in $D'$ is a property
colored cycle in $D$.\qed

\2

\begin{corollary}
For every fixed $c\ge 2$, we have $d(n,c)=\Theta(\log_2 n)$ and
$\vec{d}(n,c)=\Theta(\log_2 n)$.
\end{corollary}

\section{Open Problems}

We believe that there are functions $s(c),r(c)$ dependent only on
$c$ such that $d(n,c)=s(c)\log_2 n(1+o(1))$ and
$\vec{d}(n,c)=r(c)\log_2 n(1+o(1)).$ In particular, it would be
interesting to determine $s(2)$ and $r(2).$

\end{document}